\documentclass[useAMS,usenatbib]{mn2e}
\usepackage{lmodern}

\usepackage[T1]{fontenc}
\usepackage[latin9]{inputenc}
\usepackage{color}
\usepackage{amsmath}
\usepackage{graphicx}
\usepackage{amssymb}

\makeatletter

\pdfoutput=1

\usepackage[colorlinks,bookmarks]{hyperref}
\definecolor{linkblue}{rgb}{0,0,0.8}
\definecolor{linkgreen}{rgb}{0,0.5,0}
\definecolor{darkgreen}{rgb}{0,0.4,0}
\definecolor{purple}{rgb}{0.7,0.0,0.4}
\hypersetup{linkcolor=linkblue, citecolor=purple, urlcolor=linkblue}

\providecommand{\eprint}[1]{\href{http://arxiv.org/abs/#1}{#1}}
\providecommand{\adsurl}[1]{\href{#1}{ADS}}

\usepackage{ifthen}\def\eprinttmp@#1arXiv:#2 [#3]#4@{
\ifthenelse{\equal{#3}{x}}{\href{http://arxiv.org/abs/#1}{#1}
}{\href{http://arxiv.org/abs/#2}{arXiv:#2} [#3]}}
\renewcommand{\eprint}[1]{\eprinttmp@#1arXiv: [x]@}

\makeatletter
\newcommand{\vast}{\bBigg@{4}}
\newcommand{\Vast}{\bBigg@{5}}
\makeatother

\newcommand{\omegam}{\Omega_{m0}}
\newcommand{\sig}{\sigma_8}

\newcommand{\sint}{\sigma_{\rm int}}
\newcommand{\threeint}{\mu_{3,\rm int}}
\newcommand{\fourint}{\mu_{4,\rm int}}

\newcommand{\sintj}{\sigma_{{\rm int},j}}

\newcommand{\tgl}{\texttt{turboGL}\ }

\makeatletter

\title[First measurement of $\sigma_8$ using SN mag only]{First measurement of $\sigma_8$ using supernova magnitudes only}
\author[Castro \& Quartin]{Tiago Castro, Miguel Quartin
\\ Instituto de F\' isica, Universidade Federal do Rio de Janeiro, CEP 21941-972, Rio de Janeiro, RJ, Brazil}

\makeatother

\begin{document}

\date{Accepted 2014 May 09. Received 2014 May 05; in original form 2014 March 20.}

\maketitle
\pagerange{\pageref{firstpage}--\pageref{lastpage}} \pubyear{2014}

\label{firstpage}
\begin{abstract}
    A method was recently proposed which allows the conversion of the weak-lensing effects in the type Ia supernova (SNeIa) Hubble diagram from noise into signal. Such signal is sensitive to the growth of structure in the universe, and in particular can be used as a measurement of $\sigma_8$ independently from more traditional methods such as those based on the CMB, cosmic shear or cluster abundance. We extend here that analysis to allow for intrinsic non-Gaussianities in the supernova PDF, and discuss how this can be best modelled using the Bayes Factor. Although it was shown that a precise measurement of $\sigma_8$ requires $\sim10^5$ SNeIa, current data already allows an important proof of principle. In particular we make use of the 706 supernovae with $z\le 0.9$ of the recent JLA catalog and show that a simple treatment of intrinsic non-Gaussianities with a couple of nuisance parameters is enough for our method to yield the values $\,\sigma_8 =  0.84^{+0.28}_{-0.65}\,$ or $\,\sigma_8 < 1.45\,$ at a $2\sigma$ confidence level. This result is consistent with mock simulations and it is also in agreement with independent measurements  and presents the first ever measurement of $\sigma_8$ using SNeIa magnitudes alone.
\end{abstract}



\maketitle

\section{Introduction}\label{sec:intro}

Type Ia supernovae (SNeIa) are arguably the most important and reliable estimators of extragalactic distances. As it is well know, they provided the first solid evidence of the present cosmological acceleration~\citep{Riess:1998cb,Perlmutter:1998np}. Since this discovery a large effort has been devoted to testing and improving the calibration of the SNeIa and to correcting their light curves in order to understand and control systematics~\citep{Kessler:2009ys,Conley:2011ku,Betoule:2012an,Scolnic:2013aya}.

As their light comes from high redshifts (up to $z\simeq2$) gravitational lensing from intervening matter is expected to play an important role. The correction induced by lensing will in fact become a major source of uncertainty when richer and deeper SNeIa catalogs are compiled in the next years. The Large Synoptic Survey Telescope (LSST) project plans for instance to collect over a million SNeIa in ten years~\citep{Abell:2009aa}, roughly a thousand-fold increase from number of SNeIa observed so far. A great effort is therefore being put forward to better understand this and avoid biases; see e.g.~\citep{Jonsson2008,Amendola:2010ub,Takahashi:2011qd,Clarkson:2011br,Bolejko:2012ue,BenDayan:2013gc,Zitrin:2013jza}.

Gravitational lensing changes the intrinsic distribution function of the SNeIa magnitudes, increasing the scatter and introducing non-Gaussianity.
In~\citep{Amendola:2013twa}, we have obtained the lensing variance, skewness and kurtosis of the SNeIa distribution via sGL, a fast simulation method developed in~\citep{Kainulainen:2009dw,Kainulainen:2010at,Kainulainen:2011zx}. When confronted to $N$-body simulations sGL was shown to be very accurate up to $z \simeq 1.5$, with the advantage of results being given as  function of the relevant cosmological parameters. They also were in very good agreement with observational data~\citep{Jonsson:2009jp,Kronborg:2010uj,Jonsson:2010wx} and with other recent independent theoretical estimations~\citep{BenDayan:2013gc}. These fits can be employed to take into account the lensing extra scatter for any value of the cosmological parameters and also to model the lensing non-Gaussianity. This fact was explored in~\citep{Quartin:2013moa} where we proposed to use these accurate determinations of the lensing moments to  measure cosmological parameters, following the ideas first discussed in~\citep{Bernardeau:1996un,Hamana:1999rk,Valageas:1999ir} and later further developed in~\citep{Dodelson:2005zt}. We showed that by using not just the variance of the lensing signal but the third and fourth order moments as well, a more precise and robust measurement was possible. In a $\Lambda$CDM scenario it was verified that the most sensitive cosmological parameters to supernova lensing were $\Omega_{m0}$ and $\sigma_8$. Now since the former is already tightly constrained by the measurement of the supernova magnitudes themselves (i.e., by the first moment of the distribution), the most important \emph{new} information gained was that pertaining to $\sigma_8$.

In particular it was shown that $\sigma_8$ could be measured by the LSST survey to within 3--7\%, a value that is competitive with usual methods based on cosmic shear, cosmic microwave background (CMB) or cluster abundance, and completely independent of these. In particular, it does not rely on measuring galaxy shapes
and is thus immune to the systematics associated to the cross-correlation of intrinsic galaxy ellipticities. Also, it does not require to extrapolate the amplitude $\sigma_8$ from recombination epoch to today, as with the CMB technique, nor to make assumptions on the threshold of formation of structures that is needed when employing galaxy clusters. It also complements the method 
proposed in~\citep{Gordon:2007zw}, to wit correlating nearby supernova magnitudes with their positions to obtain their peculiar velocity correlations, which is also sensitive to $\sigma_8$.

Here we extend on previous works on two fronts. First, we generalize the method to include intrinsic non-Gaussianities in the SNeIa distributions (that is, excluding all lensing effects). We do so by employing one nuisance parameter for each central moment of the distribution. We then argue that this is the most straightforward extension of the standard supernova analysis and that a more complicated parametrization should only be used if data itself demands it; the Bayes Factor is a nice and simple way to decide which parametrization to use. Second, we apply the above generalized procedure to two real supernova catalogs: the recently published combined SDSS-II and SNLS 3-year results~\citep{Betoule:2014frx}, dubbed the Joint Lightcurve Analysis (JLA) catalog and the older standard SNLS 3-year catalog (SNLS3)~\citep{Conley:2011ku}.  We find that the method works as is, even though data is usually not treated for systematics that affect the higher moments. We thus obtain the first measurement of $\sigma_8$ from supernova magnitudes alone.

This letter is organized as follows. In Section~\ref{sec:moments} we summarize our methodology. In Section~\ref{sec:bayes}  we   show how the Bayed Factor can be used to best model the SNeIa probability distribution function (PDF), and in Section~\ref{sec:snls} we apply our method to real data. Finally, we conclude in Section \ref{sec:conclusions}.

\section{The Method of the Moments}\label{sec:moments}

Here we summarize the main point of the \emph{method-of-the-moments} (MeMo), originally discussed in~\citep{Quartin:2013moa}. In a nutshell, the idea is to use the scatter in the Hubble diagram to measure $\{ \omegam, \sig \}$ by measuring the mean $\mu_{1}'$ and the first three central moments (which we will collectively refer to simply as $\mu_{1-4}$). The moments of the lensing PDF $\mu_{1-4,\rm lens}$ were originally obtained from \href{http://www.turbogl.org/}{\tgl}and accurate fitting functions were made available in~\citep{Amendola:2013twa}. They are related to the full (observed) central moments $\mu_{1-4}$ by
\begin{align}
    \mu_{2} & \;\equiv\;\sigma_{{\rm tot}}^{2}
    \;=\;\sigma_{{\rm lens}}^{2}+\sigma_{\rm int}^{2}\,,\label{eq:mu2}\\
    \mu_{3} & \;=\;\mu_{3,{\rm lens}} + \threeint \,,\label{eq:mu3}\\
    \mu_{4} & \;=\;\mu_{4,{\rm lens}}+6\,\sigma_{{\rm lens}}^{2}\,\sigma_{\rm int}^{2} +  3 \, \sigma_{\rm int}^{4} + \fourint \,,\label{eq:mu4}
\end{align}
where $\{\sigma_{\rm int},\threeint,\fourint\}$ are the ``intrinsic'' SNeIa dispersions, which we define including any experimental contributions. The number of moments to be used in this analysis is in principle arbitrary as each new moment adds information. However, it was shown in~\citep{Quartin:2013moa} that for supernovae almost all of the information is already included using $\mu_{1-4}$ (and a very good fraction of it already in $\mu_{1-3}$).

The MeMo likelihood at each redshift bin is obtained directly from the first four moments $\mu_{1-4}$:
\begin{align}
    &L_{\rm MeMo}(\omegam, \sig, \{\sintj\}) = \exp \bigg( - \frac{1}{2} \sum_{j}^{{\rm bins}} \chi_{j}^2 \bigg) \,,\label{Lmom} \\
    &\chi^2_j = \big(\boldsymbol{\mu}-\boldsymbol{\mu}_{\rm data}\big)^t \;\Sigma_j^{-1}\; \big(\boldsymbol{\mu}-\boldsymbol{\mu}_{\rm data}\big) \,, \label{chi2mom} \\
    &\boldsymbol{\mu} = \{ \mu_1',\,\mu_2,\,\mu_3,\,\mu_4 \} \,,
\end{align}
where the vector $\boldsymbol{\mu}(z_{j},\sigma_{8},\Omega_{m0}, \sint)$ is the theoretical prediction for the moments, and its second-to-fourth components are defined in~\eqref{eq:mu2}--\eqref{eq:mu4}. The mean $\mu_{1}'$ is the  theoretical distance modulus.
The quantity $\boldsymbol{\mu}_{\rm data}(z_{j})$ is the vector of fiducial or measured (sample) moments. In forecasts  it is $\boldsymbol{\mu}(z_{j},\sigma_{8},\Omega_{m0}, \sint)$ evaluated at the fiducial model, while for real data it is best to use unbiased estimators of the central moments (sometimes called $h$-statistics, see~\citep{dwyer1937}). For instance for the third moment
\begin{align}
    \mu_{3,{\rm data}}(z_{j}) & = \sum_{k} N_j \frac{ \big[m_{k,j}-\mu_{1,{\rm data}}'(z_{j}) \big]^3}{(N_j-1)(N_j-2)} \,, \label{sample3}
\end{align}
where $m_{k,j}$ are the SNeIa distance moduli observed in the redshift bin centered at $z_{j}$. The covariance matrix $\Sigma$ is built using the fiducial (or observed) moments and therefore does not depend explicitly on cosmology (but it does on $z$).  The full covariance matrix for $\mu_{1-4}$, which appears in~\eqref{chi2mom}, can be found in~\citep{Quartin:2013moa}.

Note that the estimators found in~\citep{Quartin:2013moa} are in fact biased estimators, which only converge to the unbiased ones in the limit of large number of data points in each bin. For forecasts, such as the ones carried out in~\citep{Quartin:2013moa} this is irrelevant, but for real data here employed we find small but non-negligible corrections due to the fact that most bins have less than 50 SNeIa. Note that for such a small number of data points there are also small corrections to the full covariance matrix
, the computation of which is straightforward using computer algebra software (we employed the \textsc{Mathematica} package \textsc{MathStatica}) but the result is too large to present here explicitly.

\begin{figure*}
    \begin{centering}
    \includegraphics[width=1.5\columnwidth]{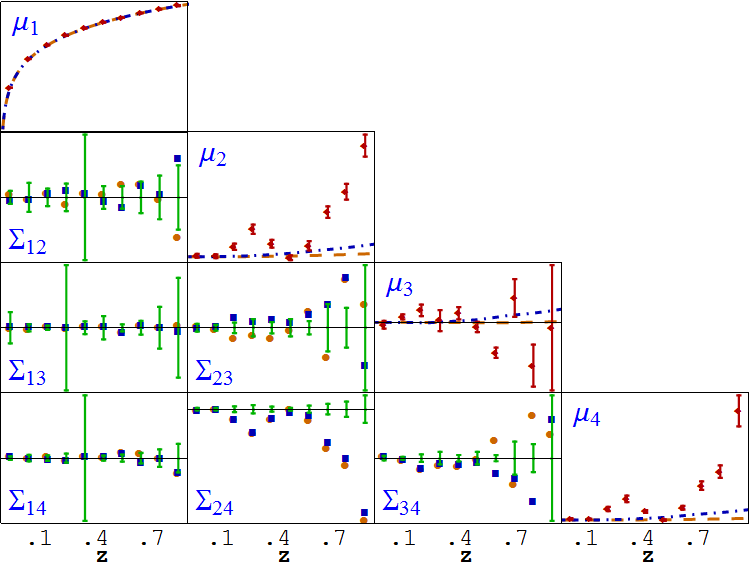}
    \caption{All 10 independent terms in the MeMo likelihood. The diagonal plots depict the measured central momenta together with the weak lensing prediction for 2 values of $\sigma_8$: the CMB fiducial ($\sigma_8=0.8$, dashed orange) and $\sigma_8=1.6$ (dot-dashed blue), which we exclude at $2.3\sigma$. In the off-diagonal cases data and model intermix, so instead we plot $\Sigma_{XY}$:
    points above (below) zero increase (decrease) the $\chi^2$.
    Although in some cases the modelling of the intrinsic non-Gaussianities as extra moments constant in $z$ (using 2 nuisance parameters) looks simplistic, the get $\,\chi^2$/d.o.f. = $1.3$. However, the last bin is an outlier, so we remove it and get a very good $\,\chi^2$/d.o.f. = $1.06$.
    \label{fig:snlscov}}
    \end{centering}
\end{figure*}

\section{Dealing with the Intrinsic Supernova PDF with the Bayes Factor}\label{sec:bayes}

\begin{table*}
    \centering
		\caption{Bayesian model comparison between different hypothesis on $\sint$ and $\threeint$}
    \centering
    \begin{tabular}{l*{5}{c}r}
    \hline\hline
    \multicolumn{3}{c}{Hypothesis} &&Probability \\
    \cline{1-3}
    Data & Model 1 & Model 2 & $\;\;\;\ln B_{12}\;\;\;$ & of best model  & $\;\sigma-$level$\;$ \\
    \hline
    $\mu_{1-2}$ (JLA) & const. & $\sint(z)$ & $-47$ &$ 1 - 4\times 10^{-21}$ & 9.4\\
    $\mu_{1-4}$ (JLA) & const. & $\sint(z)$ \& $\threeint(z)$ & 60 &$1-9\times10^{-27}$ & 10.7\\
    $\mu_{1-4}$ (JLA) & fixed in best fit & const. & 11 &$1-2\times10^{-5}$ &4.3\\
    \hline
    $\mu_{1-4}$ (DES) & const. & $\sint(z)$ \& $\threeint(z)$ & 190 &$1-3 \times 10^{-83}$&19\\
    \hline
    \hline
    \end{tabular}
	\label{tab:bayes}
\end{table*}

When the MeMo was originally proposed in~\citep{Quartin:2013moa} the assumptions made about the intrinsic supernovae dispersion was at the same time both conservative and aggressive. The SNeIa were allowed to have a dispersion which in one hand had a different $\sigma_{\rm int}(z)$ in each redshift bin, but in the other 
was assumed to be Gaussian in each bin. Real SNeIa data may nevertheless contain non-Gaussianities which are not due to lensing, either intrinsic or due to systematics and/or to the lightcurve fitting procedures.

Here we generalize the method to include non-Gaussianities in the form of intrinsic third and fourth central moments. However, if we allowed all three parameters to be free in every bin we would have no less than 30 nuisance parameters to marginalize over! Clearly this is too conservative, and instead we can do much better by following the same prescription used for the standard supernova analysis, which uses only $\mu_1'$. In that case, the supernova give the distance modulus up to a single nuisance parameter $M$, which describes the intrinsic magnitude of the supernovae, and which is assumed to be constant in $z$. In fact, a fine tuned $M(z)$ is able to fit all supernova data without any need for a cosmological constant or accelerated expansion. Clearly this is a contrived scenario, and cosmologist find it best to keep $M$ as a constant parameter and interpret supernovae data as an indication of cosmic acceleration. The same approach is probably best also for lensing, and we should only go beyond constant $\sint$, $\mu_{3,{\rm int}}$ and $\mu_{4,{\rm int}}$  if data demands it. In fact, for both catalogs here employed $\mu_{4,{\rm int}} = 0$ was either the preferred value or very close to it, so for simplicity henceforth we assume, unless otherwise stated, that $\mu_{4,{\rm int}} = 0$. This has only a small effect on the end results.

The best way to decide whether additional nuisance parameters are necessary is through the Bayes Factor ($B_{12}$)~\citep{Trotta:2005ar,Trotta:2008qt,Verdinelli95}, which is just a ratio of the so-called ``evidences'' of two models. The evidence is just the integral of the posterior over all data, and is usually neglected in parameter estimations. It is nevertheless very useful to compare models because it not only prefers models that fit best the data but has also a built-in ``Occam's Razor'' property. It is usually employed in conjunction with the Jeffrey's scale to decide which model is best. Here we went further and converted probabilities, given by $1 /(1+\exp|B_{12}|)$, into $\sigma-$levels assuming Gaussian errors (i.e., $\,0.32 \rightarrow 1\sigma$, $\,0.05 \rightarrow 2\sigma$, $\,0.003 \rightarrow 3\sigma$ and so forth). We believe this makes it simpler to interpret the results. We thus computed $B_{12}$ for real data 
in order to decide which is the best way to parametrize the intrinsic dispersion of the SNeIa. We conclude that a constant $\sint$ and $\threeint$ is favored over $\sint(z)$ and $\threeint(z)$. The results are in table~\ref{tab:bayes}. For future data from the Dark Energy Survey (DES), we did a similar test this time assuming a constant $\sint$ and $\threeint$ as fiducial. The results  clearly show that if that is the case, data will strongly favor the simpler model. It is possible that more complex modelling of intrinsic non-Gaussianity will be needed in the future for very large catalogs such as the one from LSST~\citep{Abell:2009aa}, but this can be tested as above.

We also tested the MeMo for the Union 2.1 catalog~\citep{Suzuki:2011hu}. However, we found that for the complete catalog we could not get a good fit (too high  $\chi^2$/d.o.f.). This may be due to the fact that it is a compilation of SNeIa from many different surveys.  Although care was taken to homogenize the catalog (and that a recent blind search for systematics in~\citep{Amendola:2012wc,Heneka:2013hka} found no evidence of any), the focus has always been on $\mu_1'$, whereas here the lensing signal comes from higher moments.

\begin{figure*}
\begin{centering}
    \includegraphics[width=0.93\columnwidth]{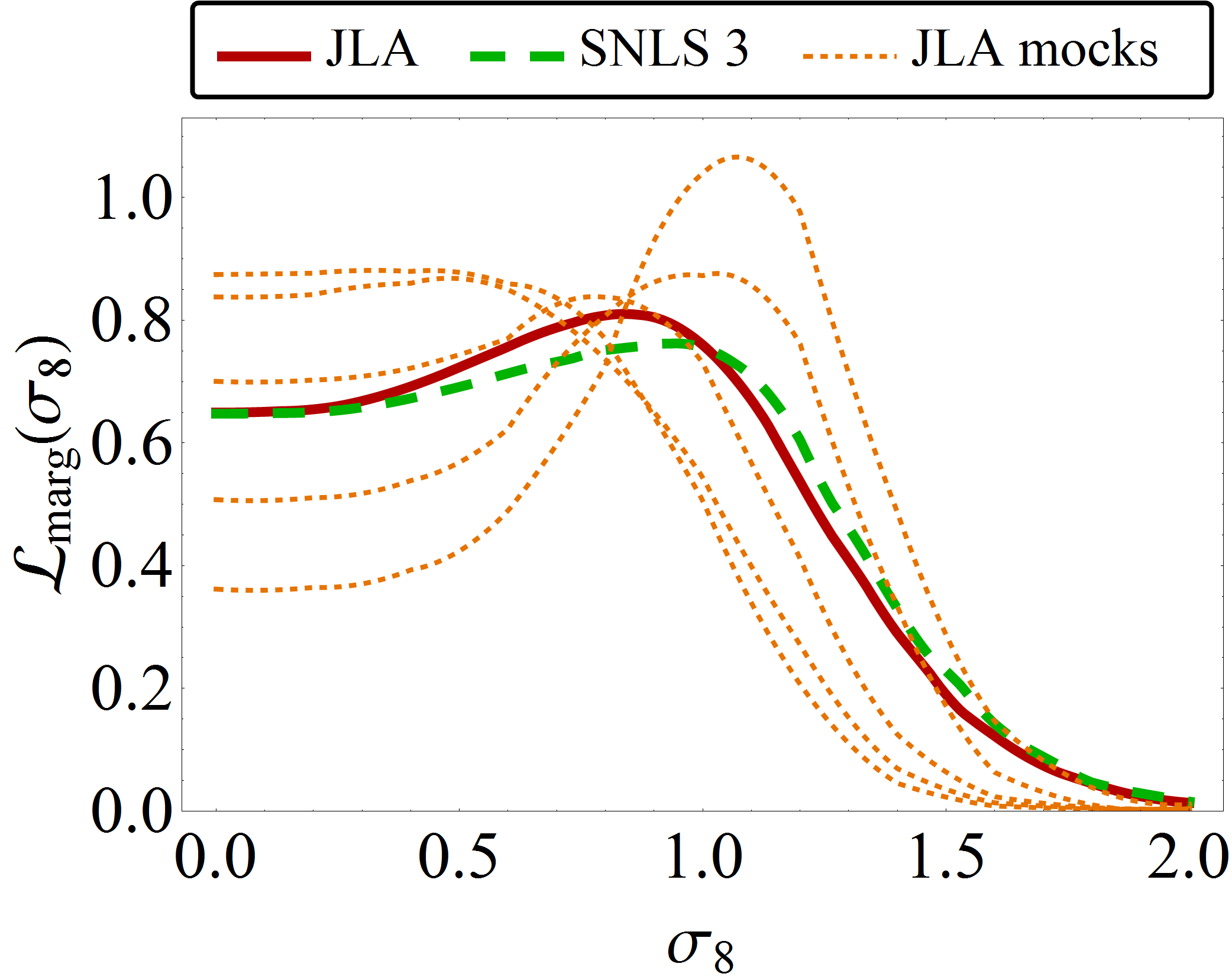}
    \qquad\qquad
    \includegraphics[width=0.93\columnwidth]{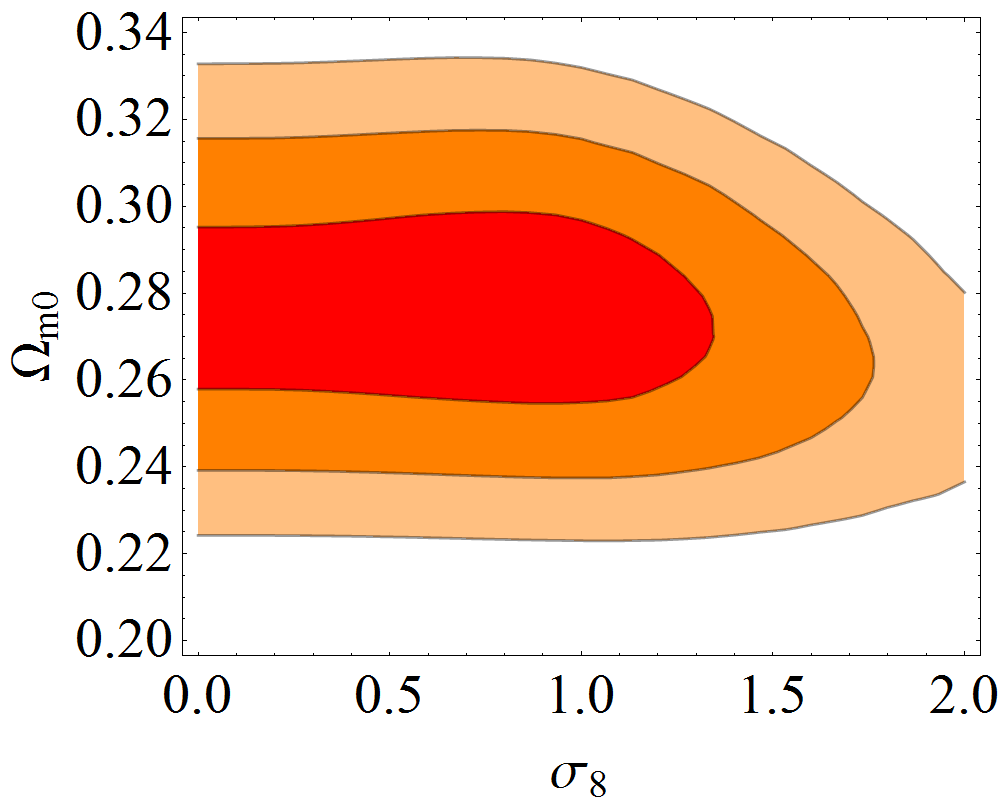}
    \caption{\emph{Left:} Posteriors on $\sigma_8$ from JLA and SNLS3 data after marginalizing over all other parameters.      We assume that both $\sint$ and $\threeint$ are constant in redshift and that $\fourint = 0$. The solid dark red curve is the posterior using the real JLA data; the green long-dashed curve is the same for SNLS 3-year data; the orange dotted curves are 5 different forecasts using mock catalogs with the same number and redshift distribution of SNeIa as the JLA catalog. \emph{Right:} Same for $\{\sigma_8,\Omega_{m0}\}$ for JLA.
    \label{fig:snls}}
\end{centering}
\end{figure*}

\section{Measuring $\sigma_8$ with JLA and SNLS3 supernova catalogs}\label{sec:snls}

In this Section we apply the method for data with $z \le 1$ in two supernova catalogs: SNLS3 (460 supernovae) and JLA (732 supernovae). The reason for the cutoff at $z=1$ is that both catalogs have too few supernovae beyond that, making it pointless and error prone any attempt to compute the central moments in that range. We employ a simple binning of the data in 10 redshift bins of 0.1 width. Since the distance modulus change inside each bin is significant, care must be taken when computing the central moments. One cannot use $m_{k,j}$ in~\eqref{sample3} directly as the measured distance moduli of each supernova. Instead, $m_{k,j}$ should be evaluated as the distance modulus at $z_j$ at the bin center plus the deviation $\Delta m_{k,j}$ with relation to the best fit curve  $m_{\rm best}(z)$. In other words:
\begin{align}
     m_{k,j}  \;\equiv\; m_{k,z_k}^{\rm catalog} - m_{\rm best}(z_k) + m_{\rm best}(z_j) \,.
\end{align}
Moreover, since current data does not put tight constraints in $\sigma_8$, we extended the numerical simulations in~\citep{Amendola:2013twa} for a broader range of values, namely $0<\sigma_8<2$.

Figure~\ref{fig:snlscov} depicts all 10 central moment terms in the likelihood, together with the expectation due to lensing assuming two different values of $\sigma_8$.

Figure~\ref{fig:snls} [left] depicts the marginalized posterior of $\sigma_8$ for the JLA and SNLS3 data, together with 5 mock catalogs with the same number and redshift distribution of SNeIa as the JLA catalog. Figure~\ref{fig:snls} [right] shows the marginalized posterior of $\{\sigma_8, \Omega_{m0}\}$ for the JLA catalog. For JLA the last $z$ bin (with only 26 SNeIa) is an outlier, so we removed it. We then get $\,\sigma_8 = 0.84^{+0.28}_{-0.65} \,$ or that $\,\sigma_8 < 1.45\,$ at a $2\sigma$ confidence level. The overall $\,\chi^2$/d.o.f. is a very good $1.06$ (if we kept the last bin, $\,\chi^2$/d.o.f. = $1.3$).
For the mock catalogs we use as fiducial values for the moments of the intrinsic SNeIa PDF the values obtained in the best-fit of the JLA catalog. It is interesting to note that even for the older SNLS3 catalog one can gets $\,\sigma_8 = 0.93^{+0.24}_{-0.72} \,$ or that $\,\sigma_8 < 1.49$. This is the first time information on cosmological perturbations is obtained from SNeIa data alone.

\begin{figure}
\begin{centering}
    \includegraphics[width=.9\columnwidth]{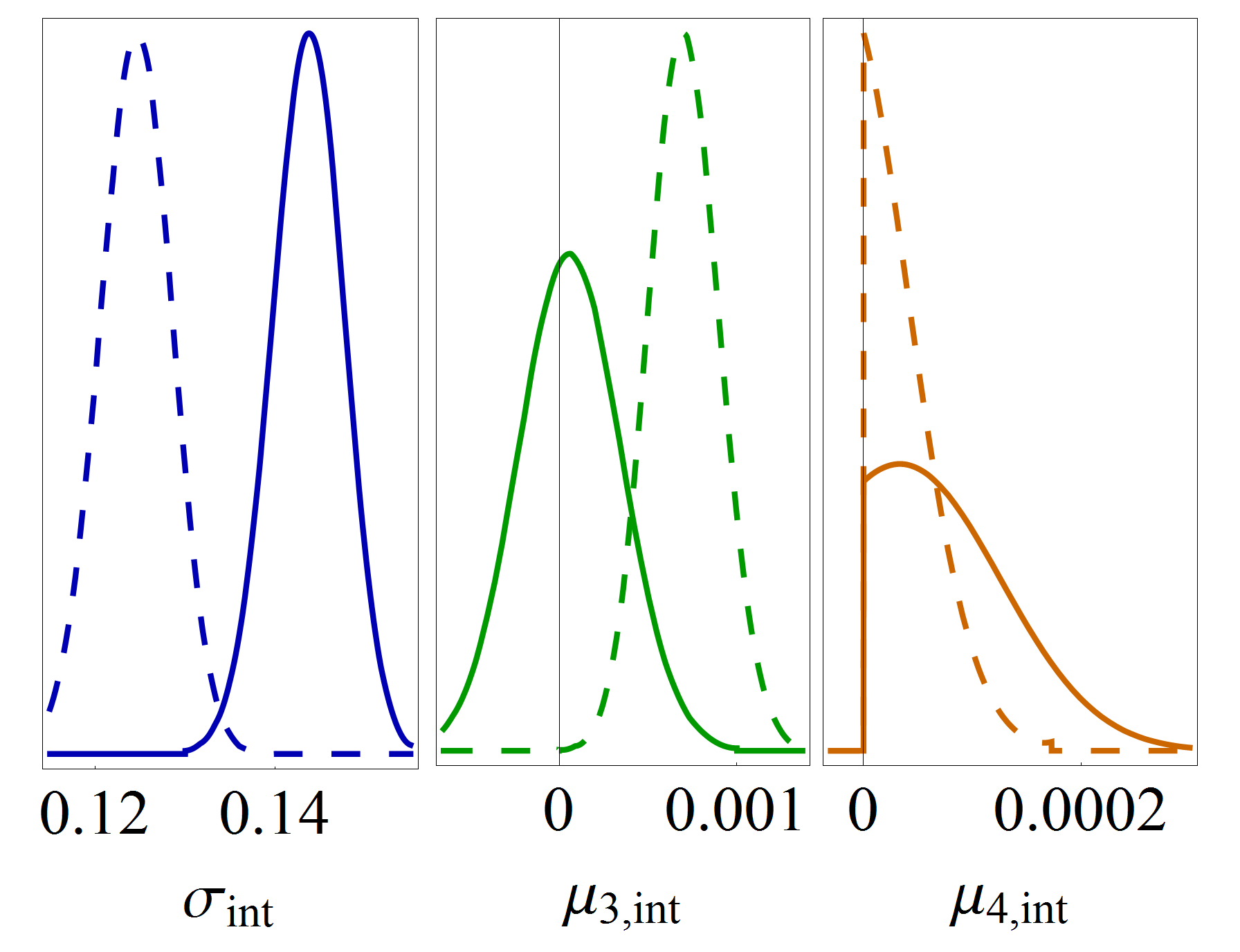}
    \caption{Intrinsic moments, in magnitudes, for the JLA (solid curves) and SNLS3 (dashed) catalogs. For $\mu_3$, JLA accepts $\mu_{3,{\rm int}} = 0$, while for SNLS $\,\mu_{3,{\rm int}} =  (6.1\pm1.9) \times 10^{-4}$. Note that in both catalogs $\mu_{4,{\rm int}} = 0$ is well inside $1\sigma$.  \label{fig:mu-int}}
\end{centering}
\end{figure}

Figure~\ref{fig:mu-int} shows the marginalized likelihoods  for the intrinsic moments (our nuisance parameters). In both catalogs $\fourint = 0$ is well inside $1\sigma$. For $\mu_3$, for JLA one has $\mu_{3,{\rm int}} = (0.8\pm2.7) \times 10^{-4}$, while for SNLS3 we find $\mu_{3,{\rm int}} =  (6.1\pm1.9) \times 10^{-4}$.

In table \ref{tab:bayesmock} we compute the evidence for lensing in JLA, SNLS3 and future surveys in detecting lensing signal. We find that JLA can only give a very faint hint at the existence of lensing ($0.9\sigma$), and even that only when using all  4 moments. In fact, using only the variance as usually done in the literature, this faint hint disappears completely, which is consistent with the results in~\citep{Karpenka:2012ys}. This is better understood in our forecasts for future DES and LSST data (using $10^5$ SNeIa) where one can clearly see that adding the third and fourth moments increases the evidence for lensing. For these forecasts we assume intrinsic Gaussianity with $\sint=0.12$ mag as our fiducial model.

\begin{table}
    \centering
    \caption{Model comparison between supernovae with lensing (model 1) and without lensing (model 2)}
    \centering
    \begin{tabular}{l*{6}{c}r}
    \hline\hline
    \multicolumn{3}{c}{Hypothesis} \\
    \cline{1-4}
    Data & $\sint$ & $\threeint$ & $\fourint$ & $\ln B_{12}$ & $\!\sigma-$lev.$\!$ \\
    \hline
    $\mu_{1-2}$ (JLA) & $\sint(z)$ & --- & --- & 0.1 & $ 0.7$\\
    $\mu_{1-4}$ (JLA) & const. & const. & $\equiv 0$ & 0.46  & 0.9\\
	$\mu_{1-4}$ (JLA) & $\sint(z)$ & $\threeint(z)$ & $\equiv 0$ & 0.17  & 0.7\\
    \hline
    $\mu_{1-2}$ (DES) & const. & -- & -- & 1.4 & 1.3\\
    $\mu_{1-3}$ (DES) & const. & const. & -- & 1.8&1.5\\
    $\mu_{1-4}$ (DES) & const. & const. & $\equiv 0$& 2.8 &1.9\\
    $\mu_{1-4}$ (DES) & $\sint(z)$ &$\threeint(z)$& $\equiv 0$ & 0.7&1.0\\
    \hline
    $\mu_{1-4}$  & const. & const. & $\equiv 0$ & 21 &6.2\\
    (LSST100k) &&&&&\\
    \hline\hline
    \end{tabular}
		\label{tab:bayesmock}
\end{table}

\section{Discussion}\label{sec:conclusions}

In this letter we obtained the first constraints for $\sigma_8$ from SNeIa data alone. In other words, without need to cross-correlate SNeIa with matter distribution data, as done for instance in~\citep{Smith:2013bha}. In order to obtain such bounds we used two nuisance parameters to cope with intrinsic scatter and skewness in the data. In principle one can use also a third nuisance parameter for the kurtosis, but data showed no need of it. In fact, for the JLA catalog even $\mu_{3,{\rm int}}$ could be set to zero, but we chose to leave it and marginalize over to get more conservative results.

Nevertheless, although the obtained bounds for $\sigma_8$ are very broad and systematics may be present, the consistency of the data with our mocks serves as an important validation of the method and opens up a new avenue in cosmology. In the future in order to best use this lensing information it is important to study whether experimental details or data reduction methods introduce systematics in the form of non-Gaussianities. Moreover, here we made use of the inferred SNeIa distances directly from JLA and SNLS3 catalogs. It would be interesting to check in detail whether including the $\sigma_8$ dependence due to lensing in the lightcurve fitter  itself  (i.e., simultaneously with the stretch and color corrections) significantly affects any of the results.

It is clear that other similar tests can be employed with our methods. For instance, one can fix completely the cosmology at, say, the CMB values and just do a hypothesis test on the data as a consistency check with lensing predictions. Other interesting possibilities would be instead to use SNeIa lensing to test either the power spectrum directly~\citep{Ben-Dayan:2013eza} or the halo models~\citep{Fedeli:2013yfa}, but both require re-deriving our estimates for the central moments.

\section*{Acknowledgment}
    It is a pleasure to thank Luca Amendola, Marcos Lima, Martin Makler, Valerio Marra, Ben Metcalf, Alessio Notari and Ribamar Reis for fruitful discussions.  MQ is grateful to Brazilian research agencies CNPq and FAPERJ for support.

\bibliographystyle{mn2e_eprint}
\bibliography{cosmo-lensing}

\label{lastpage}
\bsp
\end{document}